M. Stupar · M. D. Filipović · Q. A. Parker · G. L. White · T. G. Pannuti · P. A. Jones

# A Statistical Study of Galactic SNRs using the PMN Survey

**Abstract** The Parkes-MIT-NRAO (PMN) radio survey has been used to generate a quasi all-sky study of Galactic Supernova Remnants (SNRs) at a common frequency of 4.85 GHz ($\lambda$=6 cm). We present flux densities estimated for the sample of 110 Southern Galactic SNRs (up to $\delta = -65°$) observed with the Parkes 64-m radio telescope and an additional sample of 54 from the Northern PMN (up to $\delta = +64°$) survey undertaken with the Green Bank 43-m (20 SNRs) and 91-m (34 SNRs) radio telescopes. Out of this total sample of 164 selected SNRs (representing 71% of the currently 231 known SNRs in the Green catalogue) we consider 138 to provide reliable estimates of flux density and surface brightness distribution. This sub-sample represents those SNRs which fall within carefully chosen selection criteria which minimises the effects of the known problems in establishing reliable fluxes from the PMN survey data. Our selection criteria are based on a judicious restriction of source angular size and telescope beam together with careful evaluation of fluxes on a case by case basis. Direct comparison of our new fluxes with independent literature values gives excellent overall agreement. This gives confidence in the newly derived PMN fluxes when the selection criteria are respected. We find a sharp drop off in the flux densities for Galactic SNRs beyond 4 Jy and then a fairly flat distribution from 5-9 Jy, a slight decline and a further flat distribution from 9-20 Jy though the numbers of SNR in each Jy bin are low. We also re-visit the contentious $\Sigma - D$ (radio surface brightness - SNRs diameter) relation to determine a new power law index for a sub-sample of shell type SNRs which yields $\beta = -2.2\pm 0.6$. This new evaluation of the $\Sigma - D$ relation, applied to the restricted sample, provides new distance estimates and their Galactic scale height distribution. We find a peak in the SNR distribution between 7-11 kpc with most restricted to $\pm 100$ pc Galactic scale height.

**Keywords** (ISM) supernova remnants surveys: radio (PMN)

M. Stupar
Department of Physics, Macquarie University, Sydney 2109, Australia
Tel.: +061-2-9850 8910
Fax: +061-2-9850 8115
E-mail: mstupar@physics.mq.edu.au

M. D. Filipović
University of Western Sydney, Locked Bag 1797, Penrith South, Australia

Q. A. Parker
Department of Physics, Macquarie University, Sydney 2109, Australia
Anglo-Australian Observatory, P.O. Box 296, Epping, NSW 1710, Australia

G. L. White
James Cook University, Townsville, Queensland 4811, Australia

T. G. Pannuti
Spitzer Science Center, CIT, Mailstop 220-6, Pasadena, CA 91125, USA

P. A. Jones
Australia Telescope National Facility, CSIRO, P.O. Box 76, Epping, NSW 1710 Australia

# 1 Introduction

Thanks to the new generation of optical and radio observations from Earth-based observatories and infrared, $\gamma$-ray and X-ray observations from satellites, our knowledge of supernova remnants (SNRs) is significantly improving. In particular, a large volume of current radio observations (from below 1 GHz to over 20 GHz) has enabled an improved statistical analysis of Galactic SNRs leading to their classification, determination of surface brightness, distances, physical diameters, Galactic distribution and formation (birth) rates.

One of the earliest statistical analysis of this kind comes from Ilovaisky & Lequeux (1972) where kinematic distances for several SNRs were determined. This work also applied the so-called $\Sigma - D$ relationship (a putative relation between surface brightness $\Sigma$ and diameter $D$ of a supernova remnant first mooted by Shklovsky (1960)) to provide a crude distance determination. With this they concluded that the peak in the Galactic radial distribution of the SNRs is around 3–6 kpc. In another study, Clark & Caswell (1976) used radio-continuum ob-servations at 408 MHz and 5 GHz to investigate the Galactic distribution of SNRs.

Green (1984, 1993) also discussed the overall Galactic distribution and distances to Galactic SNRs (including remnants from historical known supernova), the reliability of the $\Sigma - D$ relation and the $|z|$ dependence of $\Sigma - D$. A comprehensive summary of accumulated data for currently known Galactic SNRs is provided in the Green (2004) list of 231 objects[1]. Using estimates of radio flux density at 1 GHz interpolated from measurements and spectral indices at different frequencies, Case & Bhattacharya (1998) determined distances to Galactic shell remnants using the $\Sigma - D$ relationship. They found that the peak of the Galactic distribution is at 5 kpc from the galactic centre (with ∼40% uncertainties).

In this paper, we have used the 4.85 GHz Parkes-MIT-NRAO (PMN) radio survey to study the known population of Galactic SNRs (remnants between $-64° < \delta < +65°$). This is the only radio survey that covers (almost) the whole sky with moderate spatial resolution with the same frequency. We thus have a crucial ingredient for a proper investigation of Galactic SNRs: i.e. an all-sky ($-88° < \delta < +75°$) survey at one radio frequency.

Reasonable distance estimates for Galactic SNRs can be found for remnants where there is an optical positional coincidence with pulsars or an assumed association with H I H II regions and molecular clouds, for which independent distance estimates may also be available. Coincidence of SNR radio detections with equivalent optical observations (Green, 1984; Case & Bhattacharya, 1998 and ref. therein), where direct proper motions can be determined between long time base-line photographic observations, provides additional distance information. However, the only way to determine distances directly from radio-continuum observations for the majority of SNRs is by the $\Sigma - D$ relation (Shklovsky, 1960). The validity of this relation has yet to be universally accepted however, e.g. Green (1984). Here, we directly test this method by establishing its form via PMN fluxes for 14 shell SNR 'calibrators' for which distances have been independently estimated. We then use the newly derived relation to estimate distances for the sub-sample of 72 Galactic shell SNRs using their newly estimated surface brightness at $\Sigma_{4.85GHz}$. We discuss the significance of our results and the overall distribution of all Galactic shell SNRs.

## 2 Observations and data analysis

The PMN 4.85 GHz survey was undertaken with three different radio telescopes. The Green Bank 91-m and 43-m telescopes were used to scan the northern sky (Condon, Broderick & Seielstad, 1989, 1991). The NRAO seven-beam receiver from Green Bank was then taken to the Parkes 64-m radio telescope (Condon, Griffith & Wright, 1993) to scan the southern sky in 1990. Table 1 shows characteristics of Parkes and Green Bank radio telescopes used for PMN.

Table 1

Characteristics of the radio telescopes used for the PMN

| Telescope | FWHM (arcmin) | Max. source ext. (arcmin) | No. of SNRs detected | Ref. |
|---|---|---|---|---|
| Parkes 64-m | 4.3 | 28 | 110 | 1 |
| Green Bank 91-m | 3.5 | 22 | 20 | 2 |
| Green Bank 43-m | 7.0 | 30 | 34 | 3 |

[1]Griffith & Wright (1993); [2]Condon, Broderick & Seielstad (1989); [3]Condon, Broderick & Seielstad (1991)

The PMN survey data were taken from the two sources. The southern sky PMN data (Condon, Griffith & Wright, 1993), from declination +10° to -90°, was downloaded directly as FITS images from the Australia Telescope National Facility (ATNF) archive available online[2]. These are Parkes 64-m telescope observations. The northern sky PMN data (also known as the 87GB and 90GB surveys) data comes from the Green Bank 91-m (Condon, Broderick & Seielstad, 1989) and 43-m telescope (Condon, Broderick & Seielstad, 1991) observations. Here, FITS images were downloaded directly from the SkyView Virtual Observatory archives (skyview.gsfc.nasa.gov).

The appropriate radio data was selected for each Galactic SNR adopting the angular extents given by Green (2004). Each derived PMN image was then used to measure the flux density taking into account any nearby confusing emission (using MIRIAD; Sault & Killen 2004). Flux densities for 164 (or 71%) of the 231 Galactic SNRs catalogued by Green (2004) were estimated in this way. An additional 40 Galactic SNRs could be only partly identified so no reliable flux densities could be extracted. Some 27 SNRs listed in the Green (2004) are not found at all in the PMN survey while others have inaccurate or systematically underestimated flux densities due to a variety of reasons (e.g. angular size of source compared to telescope beam width, source contamination, blank-field adjacency effects). A brief explanation why 37 such SNRs have not been reliably detected in PMN are given below:

i) Five SNRs are not resolved with the Parkes telescope (especially around the Galactic Center) due to their small angular sizes compared to the PMN beam-width (Table 1)

---

[1] Green's list published in 2006 (at the same web page) has addition of 35 new Galactic SNRs discovered by Brogan et al. (2006). These new Galactic SNRs are mostly small in angular diameter and most probably not resolved in the PMN survey (see Table 1).

[2] ftp.atnf.csiro.au/pub/data/pmn/surveys

ii) 28 SNRs have their flux densities inaccurate due to their extension into adjacent blank fields (Table 2, Col. 11)
iii) Two SNRs have very large angular extents and their flux density estimates are unreliable
iv) Two SNRs from Table 2 are embedded in or superposed on H II regions rendering flux estimates problematic

### 2.1 Selection Criteria applied to PMN SNR sample

The PMN survey used single dish radio telescopes where the scanning has been done in elevation at or near transit. The spillover temperature (resulting from the fact that the prime focus feed of each telescope is receiving signal from the ground surrounding the telescope) varied significantly with elevation and it was necessary to subtract a "running-median baseline" in the initial PMN analysis. These subtracted baselines have reduced the flux densities of extended objects (in our case SNRs). According to Condon (2003), the east-west extension is not affected but the measured flux density of sources extended in declination will be low in the PMN. Thus, for the Parkes 64-m radio telescope observations, sources extended in declination by more than $28'$ will have low flux densities. For the Green Bank 91-m telescope, the flux densities are low if objects are over $22'$ in angular size, and finally for the 43-m telescope the flux densities are underestimated for sources over $30'$.

Table 3

Galactic SNRs in the PMN survey (PMN South, 87GB & 90GB)

| PMN SNRs | Number |
|---|---|
| Total SNRs – Green (2004) | 231 |
| Identified in the PMN | 204 (88%) |
| With an estimated flux | 164 (71%) |
| Within selection criteria | 138 (60%) |
| Published 5 GHz flux | 104 (45%) |
| $N°$ with published flux within criteria | 78 (34%) |
| $N°$ with $<10'$ extension | 38 (16%) |

We adopt these angular size limitations as the base selection criteria for this PMN SNR study. All PMN detected SNRs over $28'$, $22'$ and $30'$ extent in declination (depending on the telescope) are rejected from our study as their flux densities will be considerably underestimated (see Table 3). Consequently, of the 204 SNRs detected in the PMN data, we consider that reliable flux densities can only be estimated for 138 (60%) (for details see Table 2).

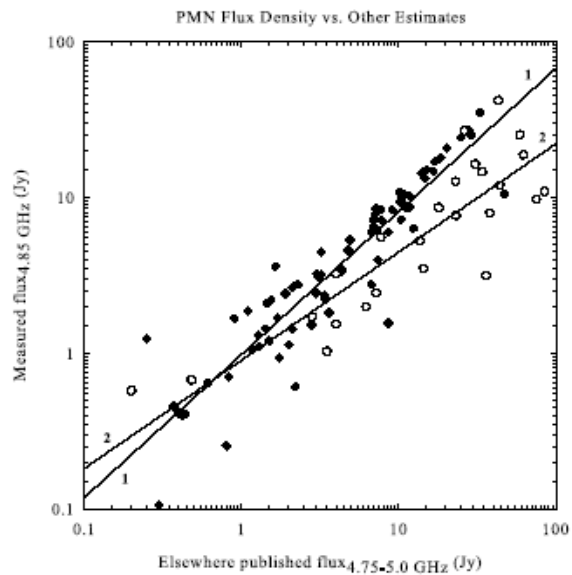

**Fig. 1** Comparison of our PMN flux densities at $S_{4.85\ \text{GHz}}$ for the 104 Galactic SNRs which also have published flux densities between 4.75 and 5.0 GHz. SNRs marked with filled diamonds are inside our source extension criteria for all three radio telescopes (see text) where the slope of the line of best fit is close to unity (0.95±0.23, trendline 1 with one outlier - refer text). Trendline 2 shows there is still a weak correlation of known and measured flux densities for objects over the extension criteria from this work (26 SNRs marked with open circle). The slope of this trendline 2 is 0.38±0.08 which not only reveals the significant underestimate of the PMN fluxes but also shows a greater scatter at larger fluxes compared to those fit by trendline 1. There is no comparison of flux densities for the two remnants G327.4+0.4 and G335.2+0.1 due to biased flux densities arising from their overlap with blank fields. G34.7-0.4 is not included in this figure (but taken into the calculation) as the flux density for this source will not fit sensibly onto the figure.

### 3 Flux Density and Surface Brightness Estimates of selected SNRs

Table 2 details the properties for all of the 164 SNRs with an estimated PMN flux at 4.85 GHz (including 26 which fall outside the selection criteria). Their measured flux densities are given in col. 5. The listed positions (RA and Dec J2000), spectral indices and diameters (in arcmin) are taken from Green (2004).

Some SNRs extend into surrounding blank fields in the PMN which probably result from solar interference affecting the receiver scanning during the survey. A few other SNRs appear "fragmented" or are mixed with noise or confused with adjacent, superposed or embedded H II regions. In such instances an accurate determination of the flux density is difficult. Such objects are indicated in Table 2 Col. 11 by an asterisk. Table 2 does not include the two very strong radio sources Cassiopeia A (G111.7-2.1) and the Crab Nebula (G184.6-5.8) as they saturate

Table 2  Properties of Galactic SNRs in PMN (4.85 GHz) Survey. Columns (4) and (6) are from Green (2004).

| (1) SNR Name | (2) RA J2000 h m s | (3) Dec J2000 ° ′ | (4) Spectral Index | (5) $S_{4.85\,\mathrm{GHz}}$ (Jy) | (6) Diam. (arcmin) | (7) $\Sigma_{4.85\,\mathrm{GHz}}$ (W m$^{-2}$ Hz$^{-1}$ sr$^{-1}$) | (8) Diam. (pc) | (9) Dist. (kpc) | (10) Dish | (11) Note | (12) Other Name(s) |
|---|---|---|---|---|---|---|---|---|---|---|---|
| G1.4-0.1 | 17 49 39 | -27 46 | ? | 2.2 | 10 | 7.8×10$^{-21}$ | 15.2 | 4.9 | 64-m | | |
| G1.9+0.3 | 17 48 45 | -27 10 | 0.7 | 0.1 | 1.2 | 1.1×10$^{-20}$ | 13.6 | 36.1 | 64-m | 1 | |
| G3.7-0.2 | 17 55 26 | -25 50 | 0.65 | 1.0 | 14×11 | 9.8×10$^{-22}$ | 12.5 | 12.2 | 64-m | | |
| G3.8+0.3 | 17 52 55 | -25 28 | ? | 0.8 | 18 | 3.7×10$^{-22}$ | 43.5 | 7.7 | 64-m | | G003.8-00.3 |
| G4.2-3.5 | 18 08 55 | -27 03 | 0.67 | 1.0 | 28 | 1.9×10$^{-22}$ | 53.5 | 6.1 | 64-m | | |
| G4.5+6.8 | 17 30 42 | -21 29 | 0.64 | 6.0 | 3 | 1.2×10$^{-19}$ | — | — | 64-m | 5 | Kepler, SN1604, 3C358 |
| G4.8+6.2 | 17 33 25 | -21 34 | 0.6 | 1.1 | 18 | 4.9×10$^{-22}$ | 38.8 | 7.4 | 64-m | | G4.5+6.2 |
| G5.2-2.6 | 18 07 30 | -25 45 | 0.44 | 1.1 | 18 | 5.5×10$^{-22}$ | 37.3 | 6.6 | 64-m | | |
| G5.4-1.2 | 18 02 10 | -24 54 | 0.27 | 16.4 | 35 | 5.4×10$^{-22}$ | — | — | 64-m | * | |
| G5.9+3.1 | 17 47 20 | -22 16 | 0.7 | 1.4 | 20 | 5.4×10$^{-22}$ | 37.6 | 6 | 64-m | | Milne56, part called G5.3-1.0 |
| G6.1+1.2 | 17 54 55 | -23 05 | 0.37 | 0.6 | 30×26 | 1.1×10$^{-22}$ | — | — | 64-m | | G6.1+1.15 |
| G7.0-0.1 | 18 01 50 | -22 54 | 0.49 | 1.4 | 15 | 9.4×10$^{-22}$ | — | — | 64-m | | G7.06-0.12 |
| G7.7-3.7 | 18 17 15 | -24 04 | 0.29 | 6.0 | 22 | 1.9×10$^{-21}$ | 24.6 | 3.6 | 64-m | | 1814-24 |
| G8.7-5.0 | 18 24 10 | -23 48 | 0.3 | 3.4 | 26 | 7.7×10$^{-22}$ | 33.3 | 4.1 | 64-m | | |
| G8.7-0.1 | 18 05 30 | -21 26 | 0.5 | 18.7 | 45 | — | — | — | 64-m | * | (W30) |
| G9.8+0.6 | 18 05 08 | -20 14 | 0.5 | 0.8 | 12 | 8.0×10$^{-22}$ | 32.9 | 8.8 | 64-m | | |
| G10.0-0.3 | 18 08 39 | -20 25 | 0.8 | 0.7 | 8 | 1.6×10$^{-21}$ | — | — | 64-m | 1 | |
| G11.2-0.3 | 18 11 27 | -19 25 | 0.49 | 9.3 | 4 | 8.8×10$^{-20}$ | — | — | 64-m | 1 | |
| G11.4-0.1 | 18 10 47 | -19 05 | 0.5 | 2.4 | 8 | 9.5×10$^{-21}$ | 14.3 | 5.7 | 64-m | | |
| G12.0-0.1 | 18 12 11 | -18 37 | 0.7 | 1.9 | 7 | 1.2×10$^{-20}$ | — | — | 64-m | 1 | |
| G13.5+0.2 | 18 14 14 | -17 12 | 0.97 | 2.6 | 5×4 | 2.1×10$^{-20}$ | 10.9 | 8 | 64-m | | |
| G15.1-1.6 | 18 24 00 | -16 34 | 0.81 | 3.4 | 30×24 | 7.2×10$^{-22}$ | 34.1 | 4.1 | 64-m | | |
| G15.9+0.2 | 18 18 52 | -15 02 | 0.64 | 2.1 | 7×5 | 8.7×10$^{-21}$ | 14.7 | 7.8 | 64-m | | |
| G16.2-2.7 | 18 28 50 | -16 11 | 0.5 | 1.7 | 17 | 8.7×10$^{-22}$ | 32.0 | 6 | 64-m | | |
| G16.8-1.1 | 18 25 20 | -14 46 | 0.37 | 4.0 | 30×24 | 8.2×10$^{-22}$ | — | — | 64-m | | G16.85-1.05 |
| G17.4-2.3 | 18 30 55 | -14 52 | 0.73 | 0.9 | 24 | 2.4×10$^{-22}$ | 49.4 | 6.6 | 64-m | | |
| G17.8-2.6 | 18 32 50 | -14 39 | 0.37 | 1.9 | 24 | 5.0×10$^{-22}$ | 38.6 | 5.1 | 64-m | | |
| G18.8+0.3 | 18 23 58 | -12 23 | 0.4 | 14.8 | 17×11 | 1.4×10$^{-20}$ | — | — | 64-m | 5 | Kes 67, G18.9+0.3 |
| G18.9-1.1 | 18 29 50 | -12 58 | varies | 12.7 | 33 | — | — | — | 64-m | * | G18.95-1.1, G18.94-1.04 |
| G20.0-0.2 | 18 28 07 | -11 35 | 0.0 | 9.7 | 10 | 1.5×10$^{-20}$ | — | — | 64-m | 1 | |
| G21.5-0.9 | 18 33 33 | -10 35 | 0.0 | 7.2 | 4 | 6.7×10$^{-20}$ | — | — | 64-m | 1 | |
| G21.8-0.6 | 18 32 45 | -10 08 | 0.5 | 25.2 | 20 | 9.5×10$^{-21}$ | 14.2 | 2.3 | 64-m | 1 | Kes 69 |
| G22.7-0.2 | 18 33 15 | -09 13 | 0.6 | 27.1 | 26 | 6.0×10$^{-21}$ | 16.7 | 2 | 64-m | 1 | |
| G23.3-0.3 | 18 34 45 | -08 48 | 0.5 | 10.5 | 27 | 2.2×10$^{-21}$ | 23.4 | 2.8 | 64-m | 1 | W41 |
| G23.6+0.3 | 18 33 03 | -08 13 | 0.3 | 5.5 | 10 | 8.2×10$^{-21}$ | — | — | 64-m | 1 | |

(Table 2, cont'd.)

| (1) SNR Name | (2) RA J2000 h m s | (3) Dec J2000 ° ′ | (4) Spectral Index | (5) $S_{4.85\text{GHz}}$ (Jy) | (6) Diam. (arcmin) | (7) $\Sigma_{4.85\text{GHz}}$ (W m$^{-2}$ Hz$^{-1}$ sr$^{-1}$) | (8) Diam. (pc) | (9) Dist. (kpc) | (10) Dish | (11) Note | (12) Other Name(s) |
|---|---|---|---|---|---|---|---|---|---|---|---|
| G24.7−0.6 | 18 38 43 | −07 32 | 0.5 | 1.8 | 15 | $1.2\times10^{-21}$ | 28.7 | 6.1 | 64-m | | |
| G24.7+0.6 | 18 34 10 | −07 05 | 0.27 | 17.1 | 30×15 | $5.8\times10^{-21}$ | – | – | 64-m | 1 | |
| G27.4+0.0 | 18 41 19 | −04 56 | 0.68 | 2.7 | 4 | $2.5\times10^{-20}$ | 10.3 | 8.2 | 64-m | | 4C−04.71, G27.3−0.1 (Kes 73) |
| G27.8+0.6 | 18 39 50 | −04 24 | varies | 8.6 | 50×30 | – | – | – | 64-m | * | |
| G28.6−0.1 | 18 43 55 | −03 53 | ? | 3.2 | 13×9 | – | 18.6 | 5.9 | 64-m | | |
| G29.6+0.1 | 18 44 52 | −02 57 | 0.5 | 0.4 | 5 | $2.4\times10^{-21}$ | 23.7 | 15.1 | 64-m | | |
| G29.7−0.3 | 18 46 25 | −02 59 | 0.7 | 2.8 | 3 | $4.6\times10^{-19}$ | – | – | 43-m | | Kes 75 |
| G30.7+1.0 | 18 44 00 | −01 32 | 0.4 | 2.2 | 24×18 | $7.6\times10^{-22}$ | 33.5 | 5.1 | 43-m | | |
| G31.5−0.6 | 18 51 10 | −01 31 | ? | 1.4 | 18 | $6.5\times10^{-21}$ | 35.7 | 6.3 | 43-m | | G31.55−0.6 |
| G31.9+0.0 | 18 49 25 | −00 55 | 0.55 | 7.2 | 7×5 | $3.1\times10^{-20}$ | – | – | 43-m | 5 | 3C391 |
| G32.8−0.1 | 18 51 25 | −00 08 | 0.27 | 8.3 | 17 | $4.3\times10^{-21}$ | 18.6 | 3.5 | 43-m | | Kes 78, K33.1−0.1 |
| G33.2−0.6 | 18 53 50 | −00 02 | varies | 0.9 | 18 | $4.1\times10^{-22}$ | 40.6 | 7.2 | 43-m | 1 | |
| G33.6+0.1 | 18 52 48 | +00 41 | 0.5 | 7.1 | 10 | $1.1\times10^{-20}$ | – | – | 43-m | 5 | Kes 79, 4C00.70, HC13, G33.7+0.0 |
| G34.7−0.4 | 18 56 00 | +01 22 | 0.30 | 115.1 | 35×27 | $1.8\times10^{-20}$ | – | – | 43-m | | W44, 3C392, G34.6−0.5 |
| G36.6−0.7 | 19 00 35 | +02 56 | ? | 0.3 | 25 | $7.2\times10^{-23}$ | 72.9 | 9.3 | 43-m | | |
| G36.6+2.6 | 18 48 49 | +04 26 | 0.57 | 0.4 | 17×13 | $2.7\times10^{-22}$ | 60.8 | 13 | 43-m | | |
| G39.2−0.3 | 19 04 08 | +05 28 | 0.6 | 10.8 | 8×6 | $3.4\times10^{-20}$ | 9.4 | 4.3 | 43-m | | 3C396, HC24, NRAO593 |
| G39.7−2.0 | 19 12 20 | +04 55 | 0.77 | 14.6 | 120 | – | – | – | 64-m | * | W50, SS433, Part called G40.0−3.1 |
| G40.5−0.5 | 19 07 10 | +06 31 | 0.5 | 2.6 | 22 | $8.1\times10^{-22}$ | 32.6 | 4.8 | 64-m | 1 | |
| G41.1−0.3 | 19 07 34 | +07 08 | 0.48 | 8.6 | 4.5×2.5 | $1.2\times10^{-19}$ | 6.1 | 5.9 | 64-m | | 3C397 |
| G42.8+0.6 | 19 07 20 | +09 05 | 0.57 | 1.2 | 24 | $3.1\times10^{-22}$ | 45.3 | 6 | 64-m | 2 | G42.8+0.65 |
| G43.3−0.2 | 19 11 08 | +09 06 | 0.48 | 18.0 | 4×3 | $2.3\times10^{-19}$ | – | – | 64-m | 5 | W49B |
| G45.7−0.4 | 19 16 25 | +11 09 | 0.47 | 1.4 | 22 | $4.4\times10^{-22}$ | 40.3 | 5.8 | 91-m | | |
| G46.8−0.3 | 19 18 10 | +12 09 | 0.5 | 7.8 | 17×13 | $5.4\times10^{-21}$ | – | – | 91-m | 5 | (HC30), G46.6−0.2 |
| G49.2−0.7 | 19 23 50 | +14 06 | 0.37 | 42.1 | 30 | – | – | – | 91-m | * | (W51) |
| G54.1+0.3 | 19 30 31 | +18 52 | 0.1 | 0.5 | 1.5 | $3.3\times10^{-20}$ | – | – | 91-m | | |
| G55.0+0.3 | 19 32 00 | +19 50 | 0.57 | 0.4 | 20×15 | $2.0\times10^{-22}$ | 54.4 | 10.1 | 91-m | | G55.2+0.5 |
| G57.2+0.8 | 19 34 59 | +21 57 | ? | 0.6 | 12 | $6.3\times10^{-22}$ | 34.9 | 9.3 | 91-m | | 4C21.53 |
| G59.5+0.1 | 19 42 33 | +23 35 | ? | 3.6 | 5 | $2.2\times10^{-20}$ | 10.7 | 6.9 | 91-m | | G59.6+0.1 |
| G59.8+1.2 | 19 38 55 | +24 19 | 0.5 | 0.6 | 20×16 | $2.8\times10^{-22}$ | – | – | 91-m | | G59.7+1.2 |
| G63.7+1.1 | 19 47 52 | +27 45 | 0.3 | 1.1 | 8 | $2.6\times10^{-21}$ | – | – | 91-m | | |
| G65.1+0.6 | 19 54 40 | +28 35 | 0.6 | 0.7 | 90 | – | – | – | 91-m | * | |
| G65.7+1.2 | 19 52 10 | +29 26 | 0.6 | 1.3 | 18 | $6.0\times10^{-22}$ | – | – | 91-m | | DA 495, G55.7+1.2 |
| G67.7+1.8 | 19 54 32 | +31 29 | 0.3 | 0.4 | 9 | $7.4\times10^{-22}$ | 33.8 | 12 | 91-m | | |
| G69.0+2.7 | 19 53 20 | +32 55 | varies | 12 | 80 | – | – | – | 91-m | * | CTB 80, G68.8+2.8 |

(Table 2, cont'd.)

| (1) SNR Name | (2) RA J2000 h m s | (3) Dec. J2000 ° ′ | (4) Spectral Index | (5) $S_{4.85\,GHz}$ (Jy) | (6) Diam. (arcmin) | (7) $\Sigma_{4.85\,GHz}$ (W m$^{-2}$ Hz$^{-1}$ sr$^{-1}$) | (8) Diam. (pc) | (9) Dist. (kpc) | (10) Dish | (11) Note | (12) Other Name(s) |
|---|---|---|---|---|---|---|---|---|---|---|---|
| G69.7+1.0 | 20 02 40 | +32 43 | 0.8 | 0.5 | 16 | $2.9\times10^{-22}$ | 49.4 | 9.9 | 91-m | | |
| G73.9+0.9 | 20 14 15 | +36 12 | 0.37 | 2.8 | 22 | $8.7\times10^{-22}$ | 32.1 | 4.7 | 91-m | | |
| G74.0−8.5 | 20 51 00 | +30 40 | varies | 10 | 230×160 | — | — | — | 91-m | * | Cygnus Loop |
| G74.9+1.2 | 20 16 02 | +37 12 | varies | 4.6 | 8×6 | $1.4\times10^{-20}$ | — | — | 91-m | | CTB 87 |
| G76.9+1.0 | 20 22 20 | +38 43 | 0.6 | 0.4 | 12×9 | $5.6\times10^{-22}$ | — | — | 91-m | | |
| G78.2+2.1 | 20 20 50 | +40 26 | 0.5 | 16 | 60 | — | — | — | 91-m | * | DR4, γ Cygni, G78.1+1.8 |
| G82.2+5.3 | 20 19 00 | +45 30 | 0.57 | 8 | 95×65 | — | — | — | 91-m | * | W63 |
| G84.2−0.8 | 20 53 20 | +43 27 | 0.5 | 1.2 | 20×16 | $5.6\times10^{-22}$ | — | — | 91-m | | |
| G84.9+0.5 | 20 50 30 | +44 53 | 0.4 | 0.4 | 6 | $1.7\times10^{-21}$ | 25.0 | 13.5 | 91-m | 4,5 | |
| G93.3+6.9 | 20 52 25 | +55 21 | 0.54 | 1.5 | 27×20 | — | — | — | 91-m | * | DA 530, 4C(T)55.38.1, G93.2+6.7 |
| G93.7−0.2 | 21 29 20 | +50 50 | 0.3 | 3 | 80 | — | — | — | 91-m | * | CTB 104A, DA 551, G93.6−0.2, G93.7−0.3 |
| G94.0+1.0 | 21 24 50 | +51 53 | 0.44 | 2.5 | 30×25 | — | — | — | 91-m | * | 3C434.1 |
| G116.9+0.2 | 23 59 10 | +62 26 | 0.57 | 1.7 | 34 | — | — | — | 91-m | * | CTB 1, G117.3+0.1, G116.9+0.1 |
| G120.1+1.4 | 00 25 18 | +64 09 | 0.61 | 24.4 | 8 | $5.7\times10^{-20}$ | — | — | 91-m | 5 | Tycho, 3C10, SN1572 |
| G127.1+0.5 | 01 28 20 | +63 10 | 0.6 | 2 | 45 | — | — | — | 91-m | * | R5, G127.3+0.7 |
| G130.7+3.1 | 02 05 41 | +64 49 | 0.10 | 35 | 9×5 | $1.3\times10^{-19}$ | — | — | 91-m | * | 3C58, SN1181 |
| G160.9+2.6 | 05 01 00 | +46 40 | 0.6 | 3 | 140×120 | — | — | — | 91-m | * | HB9, G160.5+2.8, G160.4+2.8 |
| G179.0+2.6 | 05 53 40 | +31 05 | 0.4 | 1 | 70 | — | — | — | 91-m | * | |
| G180.0−1.7 | 05 39 00 | +27 50 | varies | 3.5 | 180 | — | — | — | 91-m | * | S147 |
| G182.4+4.3 | 06 08 10 | +29 00 | 0.4 | 0.6 | 50 | — | — | — | 91-m | * | |
| G189.1+3.0 | 06 17 00 | +22 34 | 0.4 | 11 | 45 | — | — | — | 91-m | * | IC443, 3C157 |
| G260.4−3.4 | 08 22 10 | −43 00 | 0.5 | 25.3 | 60×50 | — | — | — | 64-m | * | Puppis A, MSH 08-44 |
| G284.3−1.8 | 10 18 15 | −59 00 | 0.37 | 6.5 | 24 | $1.7\times10^{-21}$ | 25.5 | 3.4 | 64-m | | MSH 10-53, G284.2−1.8 |
| G286.5−1.2 | 10 35 40 | −59 42 | ? | 0.6 | 26×6 | $5.8\times10^{-22}$ | 37.6 | 9.6 | 64-m | 1 | |
| G289.7−0.3 | 11 01 15 | −60 18 | 0.27 | 0.8 | 18×14 | $4.8\times10^{-22}$ | 40.0 | 8 | 64-m | 1 | |
| G290.1−0.8 | 11 03 05 | −60 56 | 0.4 | 20.7 | 19×14 | $1.2\times10^{-20}$ | 13.2 | 2.6 | 64-m | | MSH 11-61A |
| G291.0−0.1 | 11 11 54 | −60 38 | 0.29 | 8.3 | 15×13 | $6.4\times10^{-21}$ | — | — | 64-m | | MSH 11-62 |
| G292.0+1.8 | 11 24 36 | −59 16 | 0.4 | 10.2 | 12×8 | $1.3\times10^{-21}$ | — | — | 64-m | | MSH 11-54 |
| G292.2−0.5 | 11 19 20 | −61 28 | 0.67 | 2.5 | 20×15 | $1.3\times10^{-21}$ | 38.6 | 7.6 | 64-m | | |
| G293.8+0.6 | 11 35 00 | −60 54 | 0.67 | 1.4 | 20 | $5.3\times10^{-22}$ | — | — | 64-m | | |
| G296.5+10.0 | 12 09 40 | −52 25 | 0.5 | 7.6 | 90×65 | $1.7\times10^{-21}$ | 25.5 | 4.8 | 64-m | * | PKS 1209-51/52, G296.5+9.7 |
| G296.8−0.3 | 11 58 30 | −62 35 | 0.6 | 3.2 | 20×14 | $6.3\times10^{-21}$ | 16.0 | 5.1 | 64-m | | 1156-62 |
| G298.6−0.0 | 12 13 41 | −62 37 | 0.3 | 4.5 | 12×9 | $3.0\times10^{-22}$ | 45.8 | 10.5 | 64-m | 2 | G298.6-0.1 |
| G299.2−2.9 | 12 15 13 | −65 30 | ? | 0.4 | 18×11 | $1.8\times10^{-22}$ | 57.9 | 14.1 | 64-m | | |
| G299.6−0.5 | 12 21 45 | −63 09 | ? | 0.2 | 13 | — | — | — | 64-m | | |

(Table 2, cont'd.)

| (1) SNR Name | (2) RA J2000 h m s | (3) Dec J2000 ° ′ | (4) Spectral Index | (5) $S_{4.85\,\mathrm{GHz}}$ (Jy) | (6) Diam. (arcmin) | (7) $\Sigma_{4.85\,\mathrm{GHz}}$ (W m$^{-2}$ Hz$^{-1}$ sr$^{-1}$) | (8) Diam. (pc) | (9) Dist. (kpc) | (10) Dish | (11) Note | (12) Other Name(s) |
|---|---|---|---|---|---|---|---|---|---|---|---|
| G301.4−1.0 | 12 37 55 | −63 49 | ? | 0.5 | 27×23 | 8.8×10$^{-23}$ | 69.3 | 7.7 | 64-m | | |
| G302.3+0.7 | 12 45 55 | −62 08 | 0.47 | 2.4 | 17 | 1.3×10$^{-21}$ | 27.9 | 5.3 | 64-m | | |
| G304.6+0.1 | 13 05 59 | −62 42 | 0.5 | 6.5 | 8 | 1.5×10$^{-20}$ | − | − | 64-m | 5 | Kes 17 |
| G308.1−0.7 | 13 37 37 | −63 04 | ? | 0.2 | 13 | 1.8×10$^{-22}$ | 51.7 | 12.6 | 64-m | | |
| G309.2−0.6 | 13 46 31 | −62 54 | 0.47 | 1.1 | 15×12 | 9.2×10$^{-22}$ | 30.9 | 7.4 | 64-m | | |
| G309.8+0.0 | 13 50 30 | −62 05 | 0.5 | 3.4 | 25×19 | 1.1×10$^{-21}$ | − | − | 64-m | | G309.2-0.7 |
| G310.6−0.3 | 13 58 00 | −62 09 | ? | 4.9 | 8 | 1.2×10$^{-20}$ | 13.2 | 5.3 | 64-m | 5 | Kes 20B |
| G310.8−0.4 | 14 00 00 | −62 17 | ? | 6.5 | 12 | 6.8×10$^{-21}$ | 16.0 | 4.3 | 64-m | | Kes 20A |
| G315.4−0.3 | 14 35 55 | −60 36 | 0.4 | 5.3 | 24×13 | 2.5×10$^{-21}$ | 22.1 | 4 | 64-m | | |
| G315.9−0.0 | 14 38 25 | −60 11 | ? | 0.9 | 25×14 | 3.8×10$^{-22}$ | 41.9 | 7.1 | 64-m | | G315.8-0.0 |
| G316.3−0.0 | 14 41 30 | −60 00 | 0.4 | 13.3 | 29×14 | 5.0×10$^{-21}$ | 17.7 | 2.8 | 64-m | | MSH 14-57 |
| G317.3−0.2 | 14 49 40 | −59 46 | ? | 8.1 | 11 | 1.0×10$^{-20}$ | 14.9 | 4.1 | 64-m | | |
| G318.9+0.4 | 14 58 30 | −58 29 | 0.27 | 3.2 | 30×14 | 1.1×10$^{-21}$ | − | − | 64-m | | |
| G320.4−1.2 | 15 14 30 | −59 08 | 0.4 | 26.8 | 35 | − | − | − | 64-m | * | MSH 15-52, RCW 89 |
| G321.9−1.1 | 15 23 45 | −58 13 | ? | 1.2 | 28 | 2.3×10$^{-22}$ | 50.1 | 5.7 | 64-m | | |
| G321.9−0.3 | 15 20 40 | −57 34 | 0.3 | 3.0 | 31×23 | 6.3×10$^{-22}$ | 36.1 | 4.3 | 64-m | | |
| G322.5−0.1 | 15 23 23 | −57 06 | 0.4 | 0.6 | 15 | 4.0×10$^{-22}$ | − | − | 64-m | | |
| G323.5+0.1 | 15 28 42 | −56 21 | 0.47 | 2.0 | 13 | 1.8×10$^{-21}$ | 25.0 | 6.2 | 64-m | | |
| G327.1−1.1 | 15 54 25 | −55 09 | ? | 3.4 | 18 | 1.6×10$^{-21}$ | − | − | 64-m | | |
| G327.4+0.4 | 15 48 20 | −53 49 | 0.6 | 6.3 | 21 | 2.2×10$^{-21}$ | 23.3 | 3.6 | 64-m | 1 | Kes 27, G327.3+0.4, G327.3+0.5 |
| G327.4+1.0 | 15 46 48 | −53 20 | ? | 0.6 | 14 | 4.6×10$^{-22}$ | 41.2 | 9.4 | 64-m | | |
| G327.6+14.6 | 15 02 50 | −41 56 | 0.6 | 5.6 | − | − | − | − | 64-m | * | SN1006, PKS 1459-41 |
| G328.4+0.2 | 15 55 30 | −53 17 | 0.2 | 11.8 | 5 | 7.1×10$^{-20}$ | − | − | 64-m | 1 | (MSH 15-57) |
| G330.2+1.0 | 16 01 06 | −51 34 | 0.3 | 3.8 | 11 | 4.7×10$^{-21}$ | 18.0 | 5.2 | 64-m | | |
| G332.0+0.2 | 16 13 17 | −50 53 | 0.5 | 1.8 | 12 | 1.9×10$^{-21}$ | 24.6 | 6.6 | 64-m | 1 | |
| G332.4−0.4 | 16 17 33 | −51 02 | 0.5 | 10.4 | 10 | 1.6×10$^{-20}$ | − | 17 | 64-m | 1,5 | RCW 103 |
| G332.4+0.1 | 16 15 17 | −50 42 | 0.5 | 14.8 | 15 | 9.9×10$^{-20}$ | 14.1 | 3 | 64-m | 1 | MSH 16-51, Kes 32, G332.4+0.2 |
| G335.2+0.1 | 16 27 45 | −48 47 | 0.5 | 1.6 | 21 | 5.3×10$^{-21}$ | 37.8 | 5.7 | 64-m | 1 | |
| G336.7+0.5 | 16 32 11 | −47 19 | 0.5 | 0.6 | 14×10 | 6.5×10$^{-21}$ | 35.5 | 9.5 | 64-m | | |
| G337.0−0.1 | 16 35 57 | −47 36 | 0.67 | 15.1 | 13×7 | 6.2×10$^{-20}$ | 3.0 | 6.3 | 64-m | | (CTB 33) |
| G337.2−0.7 | 16 39 28 | −47 51 | 0.7 | 0.5 | 6 | 2.0×10$^{-21}$ | 31.2 | 17 | 64-m | 1 | |
| G337.3+1.0 | 16 32 39 | −46 36 | 0.55 | 6.2 | 15×12 | 1.0×10$^{-20}$ | 17.5 | 4.2 | 64-m | | |
| G337.8−0.1 | 16 39 01 | −46 59 | 0.5 | 8.5 | 9×6 | 2.4×10$^{-20}$ | 10.4 | 4.6 | 64-m | | Kes 40 |
| G338.1+0.4 | 16 37 59 | −46 24 | 0.4 | 0.3 | 15 | 2.0×10$^{-22}$ | 55.5 | 11.9 | 64-m | | Kes 41 |
| G338.5+0.1 | 16 41 09 | −46 19 | ? | 50.9 | 9 | 9.5×10$^{-20}$ | − | − | 64-m | 1,3 | |

(Table 2, cont'd.)

| (1) SNR Name | (2) RA J2000 h m s | (3) Dec J2000 ° ' | (4) Spectral Index | (5) $S_{4.85\,GHz}$ (Jy) | (6) Diam. (arcmin) | (7) $\Sigma_{4.85\,GHz}$ ($W\,m^{-2}\,Hz^{-1}\,sr^{-1}$) | (8) Diam. (pc) | (9) Dist. (kpc) | (10) Dish | (11) Note | (12) Other Name(s) |
|---|---|---|---|---|---|---|---|---|---|---|---|
| G340.4+0.4 | 16 46 31 | −44 39 | 0.4 | 2.9 | 10×7 | $6.2 \times 10^{-21}$ | 16.3 | 6.3 | 64-m | | |
| G340.6+0.3 | 16 47 41 | −44 34 | 0.4? | 1.5 | 6 | $6.3 \times 10^{-21}$ | 16.4 | 8.7 | 64-m | | |
| G341.2+0.9 | 16 47 35 | −43 47 | 0.6? | 1.5 | 16×22 | $6.4 \times 10^{-22}$ | — | — | 64-m | | |
| G341.9−0.3 | 16 55 01 | −44 01 | 0.5 | 1.7 | 7 | $5.2 \times 10^{-21}$ | 17.5 | 8 | 64-m | 1 | |
| G342.0−0.2 | 16 54 50 | −43 53 | 0.4? | 1.1 | 12×9 | $1.5 \times 10^{-21}$ | 26.0 | 8.1 | 64-m | 1 | |
| G342.1+0.9 | 16 50 43 | −43 04 | ? | 3.9 | 10×9 | $3.3 \times 10^{-22}$ | 68.4 | 23 | 64-m | 1 | |
| G343.1−2.3 | 17 08 00 | −44 16 | 0.5? | 2.2 | 22 | $3.2 \times 10^{-22}$ | — | — | 64-m | 3 | |
| G343.1−0.7 | 17 00 25 | −43 14 | 0.55? | 2.4 | 27×21 | $6.4 \times 10^{-22}$ | 35.5 | 4.8 | 64-m | 1 | |
| G344.7−0.1 | 17 03 51 | −41 42 | 0.5 | 1.1 | 10 | $1.7 \times 10^{-21}$ | — | — | 64-m | | |
| G346.6−0.2 | 17 10 19 | −40 11 | 0.5? | 3.4 | 8 | $8.0 \times 10^{-21}$ | 15.1 | 6 | 64-m | | |
| G348.7+0.3 | 17 13 55 | −38 11 | 0.3 | 14.4 | 17 | $7.5 \times 10^{-21}$ | — | — | 64-m | 1,5 | CTB 37B |
| G349.7+0.2 | 17 17 59 | −37 26 | 0.5 | 8.7 | 2.5×2 | $2.7 \times 10^{-19}$ | — | — | 64-m | 5 | |
| G350.0−2.0 | 17 27 50 | −38 32 | 0.4 | 5.3 | 45 | — | — | — | 64-m | * | |
| G351.2+0.1 | 17 22 27 | −36 11 | 0.4 | 3.1 | 7 | $9.5 \times 10^{-21}$ | — | — | 64-m | | G350.0−1.8 |
| G354.8−0.8 | 17 36 00 | −33 42 | ? | 0.2 | 19 | $8.3 \times 10^{-23}$ | 73.5 | 12.4 | 43-m | | G351.3+0.2 |
| G355.6−0.0 | 17 35 16 | −32 38 | ? | 0.2 | 8×6 | $1.3 \times 10^{-20}$ | 33.8 | 15.5 | 43-m | | |
| G355.9−2.5 | 17 45 53 | −33 43 | 0.5 | 2.3 | 13 | $2.0 \times 10^{-21}$ | 23.7 | 5.9 | 43-m | | |
| G356.2+4.5 | 17 19 00 | −29 40 | 0.7 | 1.3 | 25 | $3.1 \times 10^{-22}$ | 44.8 | 6.1 | 43-m | | |
| G356.3−1.5 | 17 42 35 | −32 52 | ? | 0.3 | 20×15 | $1.8 \times 10^{-22}$ | 59.3 | 11 | 43-m | | |
| G357.7−0.1 | 17 40 29 | −30 58 | 0.4 | 17.5 | 8×3 | $1.1 \times 10^{-19}$ | — | — | 43-m | | MSH 17−39 |
| G357.7+0.3 | 17 38 35 | −30 44 | 0.4? | 5.9 | 24 | $1.5 \times 10^{-21}$ | 26.6 | 3.5 | 43-m | | |
| G359.0−0.9 | 17 46 50 | −30 16 | 0.5 | 6.5 | 23 | $1.9 \times 10^{-21}$ | 24.6 | 3.4 | 43-m | | |
| G359.1−0.5 | 17 45 30 | −29 57 | 0.4? | 1.3 | 24 | $3.4 \times 10^{-22}$ | 43.5 | 5.8 | 43-m | | |
| G359.1+0.9 | 17 39 36 | −29 11 | ? | 0.4 | 12×11 | $4.5 \times 10^{-22}$ | 43.1 | 11.9 | 64-m | | |

Remarks to column 11:

\* – Flux unreliable. Source extensions outside the selection criteria
1 – Not completely seen due to blank field adjacency
2 – Not clearly seen (mixed with noise)
3 – Embedded/superposed in/on H II region
4 – Measured inside radius of ∼ 5'
5 – SNR Calibrator

the receivers and have significant sidelobes (Gregory & Condon, 1991).

The estimated uncertainties in flux densities are of the order of 10%. This estimate is based on the known uncertainties in the PMN survey (Condon, Broderick & Seielstad, 1989) as well values determined through a comparison of flux densities from other published work.

In Fig. 1 we compare our flux density estimates at 4.85 GHz for 104 Galactic SNRs with flux densities published elsewhere with similar frequencies between 4.75 GHz and 5.0 GHz (Trushkin, 2000 and references therein). A simple linear regression fit to the comparison produced a slope close to unity of 0.95±0.23 (trendline 1). This gives us confidence in our new independent PMN flux estimates.

For this trendline, only SNRs satisfying the selection criteria for all three radio telescopes were used (total of 104 out of 138). There is one outlier on this figure however (SNR G23.3−0.3) but in this case we believe the previously published flux density was poorly determined. Trendline 2 was obtained for the 26 SNRs that fall outside our selection criteria for the 43-m, 64-m and 91-m telescopes (over $30'$ $28'$ and $22'$ respectively). A weak correlation is seen but with a large scatter especially at large flux densities. The best fit line has a slope of 0.38±0.08 clearly showing the significant underestimate of the PMN flux compared to the published flux. The flux density values for trendline 2 are given in Table 2 column 5 and are marked with an asterisk.

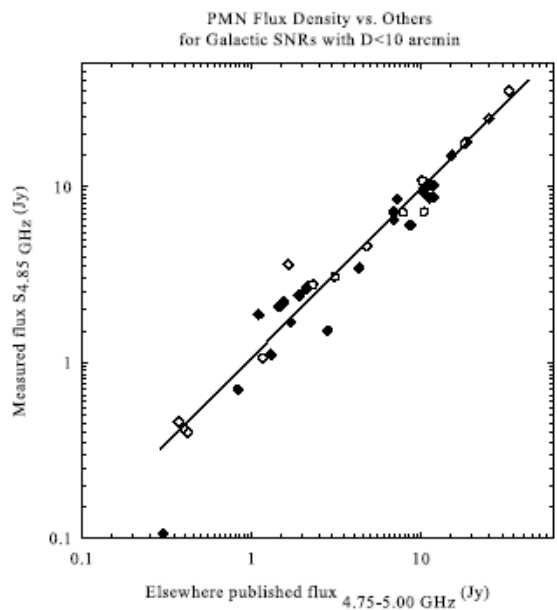

Fig. 2 Comparison of our flux densities at $S_{4.85\,GHz}$ with flux densities published elsewhere for the SNRs with diameters $\leq 10'$. The slope of line of best fit is close to unity at 0.98±0.03. Filled diamonds represent SNRs from the 64-m telescope, open circles and diamonds from the 43-m and 91-m telescopes respectively.

Obviously, the apparent scatter in the log–log plot in Fig. 1 decreases beyond 10 Janskys, at least for the SNRs within the angular size selection criteria. Note that at high flux levels (>10 Jy) the PMN fluxes diverge slightly from the unity line in the sense that the PMN fluxes are underestimated (for reasons which we have already explained in section 2). Nevertheless the tight relation indicates that a reliable correction for this can be made.

If we further tighten our selection comparison to SNRs with diameters less than $10'$ (e.g. objects with 1/3 of the baseline length for the Green Bank 91-m telescope, about 1/6 baseline the length for Parkes and 1/9 the baseline length for the 43-m telescope, refer section 2.1) the slope of the fitted regression line approaches unity with a small dispersion (0.98±0.03). Furthermore, there is no evidence of any underestimate in the PMN fluxes. These data (for 38 SNRs) are shown in Fig. 2.

We have checked the possibility that the bias in flux densities evident in Fig. 1 is flux density dependent. As shown in Figs. 3a and 3b (which plots only SNRs with diameters $\leq 10'$) no trend can be found. The SNRs flux density ratio is distributed equally about the unity line and does not appear to be either telescope or source diameter dependent.

Figure 4 gives the flux density ratio to SNR diameter dependency. Apart from the obvious and expected trend that the flux density of SNRs with larger diameters (marked with open circles in Fig. 4 and defined in section 2.1) are underestimated in the PMN, no other significant trend can be seen. Forty four percent (34 out of 78) of the Galactic SNRs (taken according selection criteria) fall within 10% of the unity flux density ratio.

There are however 28 sources outside a generous correlation of ± a factor of two. This result is not unexpected as the determination of flux densities for weak and extended radio sources is prone to uncertainties in interpretation, especially in data sets from the PMN.

3.1 Galactic SNR Flux Density distribution at 4.85 GHz

In Fig. 5 we histogram the flux density distribution for the 138 SNRs which fall within the selection criteria (which have fluxes up to 20 Jy). Most have flux densities < 4 Jy with a high concentration of almost 46% with flux densities ($S_{4.85\,GHz}$) less than 2 Jy. Beyond 4 Jy there is a sharp drop off in SNR numbers at a given flux. Between 4 and 9 Jy the distribution is essentially flat. Between 10 and 20 Jy the number of known SNRs per Jy bin drops off again by a factor of about 2 along a flat, extended high flux tail but the numbers per bin are low (0–3).

Excluded from Fig. 5 are seven SNRs with flux densities more than 20 Jy at 4.85 GHz as these will not fit sensibly onto the histogram. These are G21.8-0.6 (25.1 Jy), G22.7-0.2 (27.1 Jy), G120.1+1.4 (27.1 Jy), G290.1-0.8 (20.7 Jy), G130.7+3.1 (35 Jy), G338.5+0.1 (50.9 Jy) and

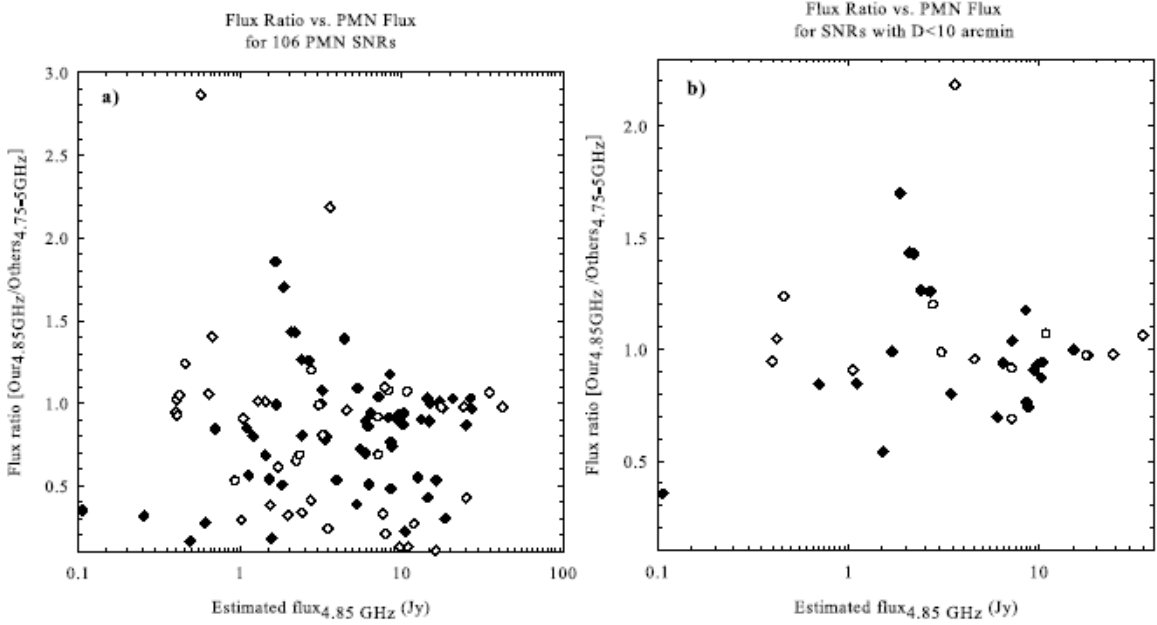

**Fig. 3** The ratio of observed flux density to flux densities estimated elsewhere for the whole sample of 104 Galactic SNRs (Fig. 3a) and for a subsample of 38 SNRs with diameter $\leq 10'$ (Fig. 3b). Filled diamonds represent SNRs from 64-m telescope, open circles from 43-m telescope and open diamonds from 91-m telescope surveys.

G34.7−0.4 which has the highest flux density in this sample at 115.1 Jy (also known as W44).

## 3.2 PMN Surface Brightness estimates of the SNR sample at 4.85 GHz

We have estimated the surface brightness at 4.85 GHz adopting the method of Clark & Caswell (1976) for the 138 Galactic SNRs which fall within our selection criteria (Table 2; Column 7) and using the PMN flux-densities from the on-line archive data together with their estimated angular diameters (in arcmin) taken from Green (2004). In the cases where the SNR boundary is not circular an equivalent diameter has been calculated according to the method suggested by Clark & Caswell (1976) and Case & Bhattacharya (1998). Our newly estimated values of surface brightness for each SNR ($\Sigma_{4.85\text{GHz}}$) are shown in Table 2, Col. 7. The majority of SNRs at this frequency are in the range $10^{-21} - 10^{-22}$ W m$^{-2}$ Hz$^{-1}$ sr$^{-1}$.

## 4 The $\Sigma - D$ relation and new distance estimates to selected Galactic SNRs

Most SNRs can be conveniently observed at radio-continuum frequencies but the number of equivalent optical detections is significantly less. This is now being at least partially addressed by the advent of sensitive new narrowband H$\alpha$ surveys such as the AAO/UKST H$\alpha$ survey (Parker et al. 2005) and SHASSA survey (Gaustad et al. 2001) of the Southern Galactic Plane and hemisphere and the IPHAS survey (Drew et al. 2005) of the Northern Galactic Plane. Many known SNRs are now yielding optical counterparts and indeed in some instances (e.g. Parker, Frew and Stupar 2004, Stupar et al. 2006) new Galactic SNR's are being found in the optical first.

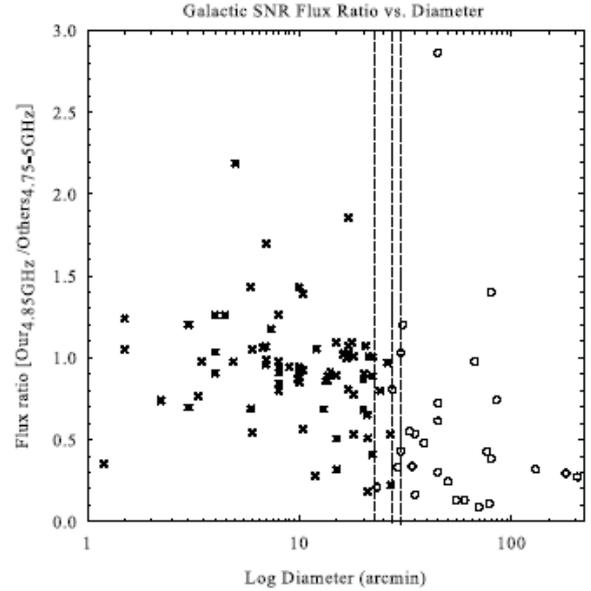

**Fig. 4** A comparison of the ratio of flux densities with estimated SNR angular diameter (this work divided by published data). Open circles represent the 28 SNRs outside of the selection criteria defined in section 2.1 and the vertical dotted lines represent the observing limitations (in arcmin) imposed for the 3 separate radio dishes according to our selection criteria. The flux efficiency of the dishes will degrade significantly once the diameter gets comparable with the FWHM of the primary beam of the dish. Therefore, some effects on the measured flux are expected to be observable when the diameter is of order 1/2 the FWHM of the primary beam.

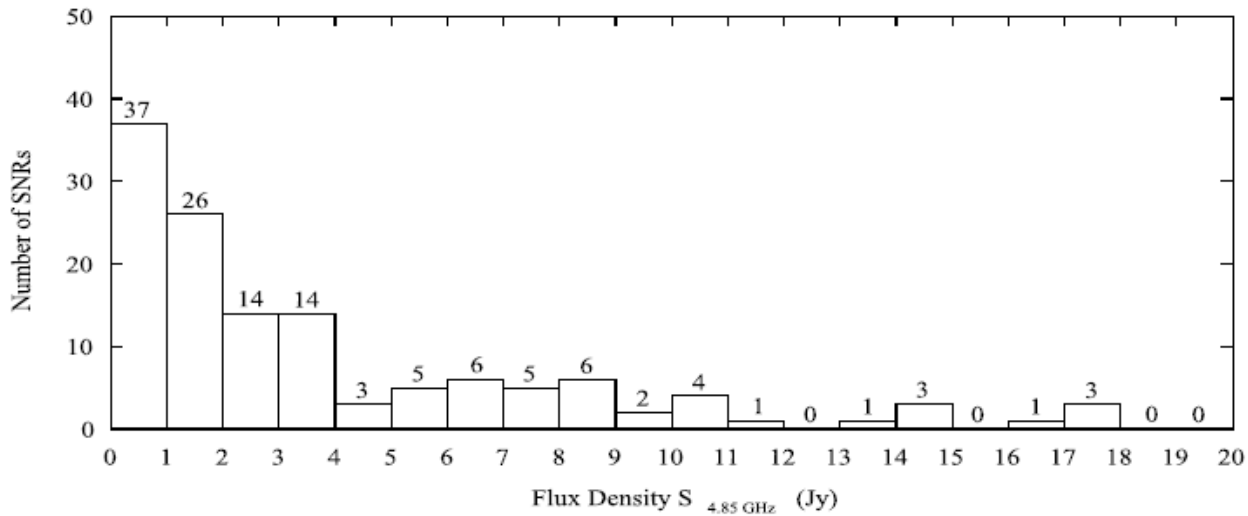

**Fig. 5** The number of Galactic SNRs (PMN South and North) as a function of flux density ($S_{4.85 GHz}$). This figure shows 131 out of our sample of 138 Galactic SNRs. Seven SNRs: G21.8-0.6 (S=25.1 Jy), G22.7-0.2 (27.1 Jy), G120.1+1.4 (24.4 Jy), G290.1-0.8 (20.7 Jy), G130.7+3.1 (35 Jy), G338.5+0.1 (50.9 Jy) and G34.7-0.4 (W44) (115.1 Jy) are not shown on this diagram.

Nevertheless, the lack of decent optical counterparts has made distance determinations for these SNRs difficult. However, when a SNR has a positional coincidence with a H II region(s), OB association, molecular cloud or a pulsar, it is possible to assume a common distance. For shell remnants with only radio observations, it has been usual, though controversial, to use the putative $\Sigma - D$ (surface brightness to diameter) relation (Case & Bhattacharya, 1998).

In principle, the surface brightness ($\Sigma_\nu$) of a SNR at a given radio frequency $\nu$ is a distance-independent parameter and is, to a first approximation, an intrinsic property of a SNR. Following Shklovsky (1960) we have the relation:

$$\Sigma_\nu = AD^\beta$$

where the distance to the SNR $d$ will be proportional to $\Sigma_\nu^{1/\beta}\theta^{-1}$. Here, $\theta$ is the SNR angular diameter, $\beta$ the power law index, an $A$ is a constant and $D$ physical diameter. To find the value of the power law index $\beta$ we use Galactic shell remnants with known distances as calibrators. The distances for these shell SNRs are determined independently, mostly via optical observations (Case & Bhattacharya, 1998 and ref. therein). Knowing these distances, the physical diameters $D$ of calibrators (in pc) can be derived.

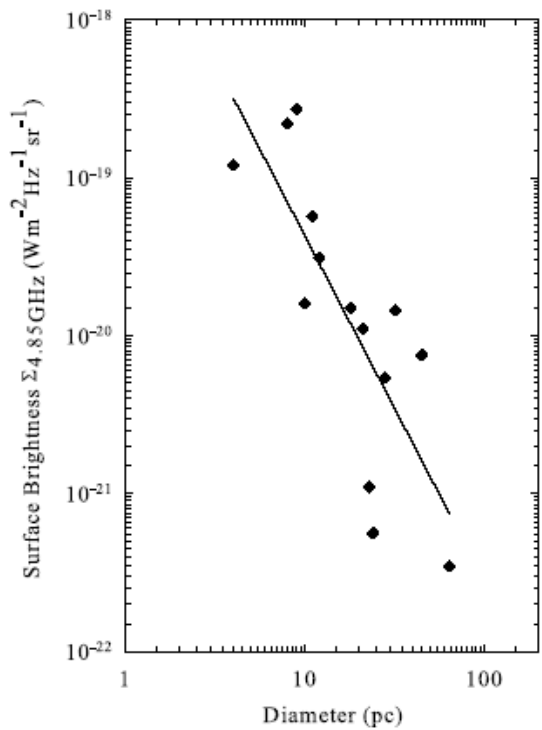

**Fig. 6** The surface brightness ($\Sigma_{4.85GHz} - D$) for 14 shell calibrators (marked in Table 2, Col. 11). A least-square fit gave a slope of $\beta = -2.2 \pm 0.6$.

Once the trend has thus been established with calibrating objects of known distance, the relation can be used for other classes of SNRs. Unfortunately there are relatively few SNRs with well determined distances to populate the relation. Consequently distances determined from the $\Sigma$–$D$ relationship may have considerable uncertainties but in the absence of other more reliable techniques it remains a useful indicator.

Thus, we have identified 14 out of 38 shell remnants as decent calibrators (marked in Table 2, Col. 11) from a recent list compiled by Case & Bhattacharya (1998). These are used to estimate the power law index $\beta$ at 4.85 GHz using their measured physical diameters arising from their known distances together with our new estimates of their surface brightness.

A least-squares fit to these Galactic SNR calibrators which fall in the range:

$$3.5 \times 10^{-22} < \Sigma_{4.85\,\mathrm{GHz}} < 2.7 \times 10^{-19}\ \mathrm{W\ m^{-2}\ Hz^{-1}\ sr^{-1}}$$

gave a slope $\beta = -2.2 \pm 0.6$.

Clark & Caswell (1976) found $\beta = -3.4$ at 5 GHz (using 20 calibrators with Cas A excluded).[3] Our new determination of the slope of the $\Sigma - D$ relation of $\beta = -2.2 \pm 0.6$ differs at the ~1.5 standard deviation level from the Clark & Caswell (1976) value. However, as we have used more recent (and more reliable) estimates of the physical diameters of the shell SNR calibrators a different value is not unexpected. Our value is also within the theoretical range determined by Duric & Seaquist (1986)[4] and based on the Sedov (1959) model for SNRs. Fig. 6 shows our $\Sigma_{4.8\,\mathrm{GHz}} - D$ sample for 14 calibrators from the Case & Bhattacharya (1998) list.

Green (1984, 1991), amongst others, has expressed skepticism regarding the use of the $\Sigma - D$ relation for any sensible SNR distance determination.[5] He concluded that the independent distance estimates which form the basis of the relation may have significant uncertainties; for example H I emission is difficult to interpret when associated with other objects such as H II regions and OB associations. Even so, Huang & Thaddeus (1985) concluded that the $\Sigma - D$ relation calibrated using shell SNRs associated with large molecular clouds shows less scatter than previous relations and does establish a good distance scale for such shell SNRs.

Using the physical diameters obtained from known distances on Fig. 6 enabled us to derive the following relation:

---

[3] Cas A has a very high surface brightness and has an extreme influence on any regression so is usually excluded from any $\Sigma - D$ relation. Likewise, in the PMN survey, Cas A saturated the receiver and so no sensible flux/surface brightness estimate is possible (see Section 3).

[4] For the theoretical work of Duric & Seaquist (1986), Case & Bhattacharya (1998) concluded that $\beta$ lies in the range $-1.9 < \beta < -5$ if typical values of SNR magnetic field evolution and radio spectral index are taken. For details see Case & Bhattacharya (1998).

[5] For application of $\Sigma - D$ on the other (nearby) galaxies see Urošević (2005).

$$\Sigma_{4.85\,\mathrm{GHz}} = 1.86^{+9.3}_{-1.6} \times 10^{-17} D^{-2.2(\pm 0.6)}\ \mathrm{W\,m^{-2}Hz^{-1}sr^{-1}}$$

This relationship was then used to estimate new distances for the 93 Galactic Shell SNRs that have not been determined by other means. Thes new estimates are given in column 9 of Table 2. Shell SNRs with known distances as well as other non-shell Galactic SNRs are excluded in this column.

We have compared our new distance estimates to these Galactic shell SNRs with those in Case & Bhattacharya (1998) which were estimated for 1 GHz. The fact that they are in good agreement shows that the flux from PMN survey can be used, not only for point like sources, but also for extended objects if established selection criteria (section 2.1) are respected.

Based on our new distance determinations for our sample of Galactic SNRs we have found their Galactic distribution (from the Sun) falls between 3 kpc and 13 kpc with a peak between 7 kpc and 11 kpc. Figure 7 shows the concentration of the Galactic SNRs as a function of Galactic longitude and height ($z$) in parsecs above and below the Galactic plane based of our new data (distances) from the Table 2. According to this figure, the furthest from the Galactic plane is SNR G4.8+6.2 at a height of 800 pc followed with G36.6+2.6 at height of 590 pc.

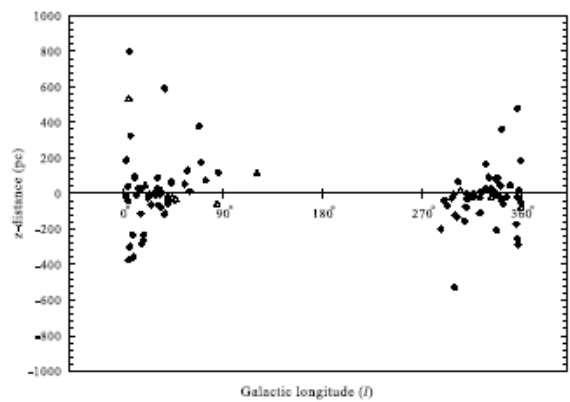

**Fig. 7** The $z$ distribution of Galactic SNRs. Data for the distances are taken from Table 2. The figure also includes the $z$ distribution of the calibrators used (marked with triangles). SNR G4.8+6.2 is at the top of the figure with a height of 800 pc above the Galactic plane.

By comparison, Ilovaisky & Lequeux (1972) found a peak in the SNR distribution at 6–8 kpc from the Sun. Kodaira (1974) used the same sample but a different empirical method for the correction of observational effects and obtained a distribution which peaked at 3–6 kpc, similar to that of Johnson & MacLeod (1963). A similar peak value of 4–6 kpc was also found by Leahy & Xinij (1989) whereas Case & Bhattacharya (1998) estimated the peak distribution at ~5 kpc with a scale length of ~7 kpc.

## 5 Conclusion

We have investigated known Galactic SNRs detected in the PMN Survey at 4.85 GHz ($\lambda$=6 cm). Using Green's (2004) list of 231 known Galactic SNRs we have determined new flux density estimates at $S_{4.85\,\mathrm{GHz}}$ for some 164 SNRs. However only 138 fall inside our conservative source diameter selection criteria established to minimize the known problems with using the PMN to derive reliable fluxes.

For these carefully selected sources we are able to demonstrate that acceptable flux densities can, if fact, be derived, together with modest uncertainties. This is confirmed by comparing the new PMN measures directly with reliable published fluxes for the same sources at similar frequencies. Indeed the relation between the PMN and published values at large flux levels is so tight that a small correction to the PMN fluxes can be made directly from the fit. The measured PMN flux density of very extended sources ($>10'$) is confirmed as being considerably underestimated as expected (recording perhaps only 30% of the true flux density when compared to other estimates) though even here a correlation with the published values is seen but with larger scatter. The PMN underestimates flux for very extended sources due to the known peculiarities of the PMN survey (refer section 2). Such sources are excluded from the subsequent analysis.

Nevertheless, we have shown that provided certain criteria are respected, PMN fluxes for Galactic shell SNRs are reliable, dispelling the myth that the PMN survey data cannot be used for such purposes. Indeed, for the sample with angular size $< 10'$, the fit with published fluxes as a function of flux yields a slope of unity with a modest scatter (Fig 2). The new PMN flux density estimates for 138 Galactic SNRs in both the Northern and Southern hemispheres revealed a sharp fall at 4 Jy and a relatively flat distribution from 4-11 Jy, a further slight drop and a flat distribution out to 20 Jy.

We have also investigated the controversial $\Sigma - D$ relationship with a view to determining new distances for our PMN sample of Galactic shell remnants currently lacking any distance estimate. To do this we first calculated the PMN surface brightness ($\Sigma_{4.85\,\mathrm{GHz}}$) for 138 SNRs. We then established a new version of the $\Sigma - D$ relation using a sample of 14 Galactic SNR calibrators whose distances have been derived independently and are considered reliable. A slope for this relation of $\beta = -2.2 \pm 0.6$ was found, in good agreement with published and theoretical predictions. This new $\Sigma - D$ was then used to obtain estimates for 93 Galactic shell remnants based on their observed angular extent and their physical diameter in parsecs. Their distribution is found to peak between 7 and 11 kpc from the Sun with most confined to within $\pm 200$ pc of the Galactic plane.

## Acknowledgments

We used the Karma and MIRIAD software packages developed by the ATNF. Many thanks to the unknown referee for excellent suggestions.